%% file: main.tex
\documentclass[a4paper,11pt]{article}
\pdfoutput=1 

\usepackage{jinstpub} 

\usepackage[utf8]{inputenc}
\usepackage{natbib}
\usepackage{graphicx}
\usepackage{svg}
\usepackage{amsmath}
\usepackage{ulem} 
\usepackage{siunitx}
\usepackage{float}
\usepackage{parcolumns}
\usepackage{qtree}
\usepackage{url}

\usepackage{color}
\usepackage{xcolor}
\usepackage{caption}
\usepackage{textcomp}
\usepackage{todonotes}
\usepackage{listings}
\usepackage{algorithm,algpseudocode}
\usepackage{comment}

\begin{document}
\input{license.tex}

\title{\boldmath A lightweight, user-configurable detector ASIC digital architecture with on-chip data compression for MHz X-ray coherent diffraction imaging}

\author[a]{Sebastian Strempfer,}
\author[b,1]{Tao Zhou\note{Corresponding authors.}, }
\author[c]{Kazutomo Yoshii,}
\author[a]{Mike Hammer,}
\author[a]{Anakha Babu,} 
\author[a]{Dawid Bycul,} 
\author[a]{John Weizeorick,} 
\author[a]{Mathew J. Cherukara,} 
\author[a,1]{Antonino Miceli}

\affiliation[a]{X-ray Science Division, Argonne National Laboratory, Lemont, IL, U.S.A.}
\affiliation[b]{Center for Nanoscale Materials, Argonne National Laboratory, Lemont, IL, U.S.A.}
\affiliation[c]{Mathematics and Computer Science, Argonne National Laboratory, Lemont, IL, U.S.A.}

\emailAdd{tzhou@anl.gov, amiceli@anl.gov}

\date{}


\abstract{Today, most  X-ray pixel detectors used at light sources transmit raw pixel data off the detector ASIC. With the availability of more advanced ASIC technology nodes for scientific application, more digital functionality from the computing domains (e.g., compression) can be integrated directly into a detector ASIC to increase data velocity. In this paper, we describe a lightweight, user-configurable detector ASIC digital architecture with on-chip compression which can be implemented in \SI{130}{\nm} technologies in a reasonable area on the ASIC periphery. In addition, we present a design to efficiently handle the variable data from the stream of parallel compressors. The architecture includes user-selectable lossy and lossless compression blocks. The impact of lossy compression algorithms is evaluated on simulated and experimental X-ray ptychography datasets. This architecture is a practical approach to increase pixel detector frame rates towards the continuous \SI{1}{\MHz} regime for not only coherent imaging techniques such as ptychography, but also for other diffraction techniques at X-ray light sources. 
}

\arxivnumber{xxxx}

\maketitle
\flushbottom

\section{Introduction}
Detectors are an integral part in scientific discovery. From Wilhelm Röntgen's barium platinocyanide screen to Georges Charpak's multiwire proportional chamber, detectors are what grant scientists the ability to see, often far beyond the human limits. Biomedical research breakthroughs such as the development of human pharmaceutical~\cite{HIV} and vaccines~\cite{covid} and the understanding of protein receptors~\cite{GPCR} have all been enabled by megapixel detectors tailored for X-ray crystallography. Through the use of high-speed direct electron counting detectors~\cite{Booth2012}, structural biologists were able to break for the first time the 0.3-nm imaging resolution barrier in cryo-electron microscopy~\cite{Li:2013gt}. More recently, using a high dynamic range pixel array detector~\cite{Tate2016}, materials scientists were able to see atoms in record breaking details (\SI{0.039}{\nano \meter} spatial resolution) with electron ptychography~\cite{Jiang:2018jna}. 

Modern pixel detectors are built on application-specific integrated circuits (ASICs) fabricated at commercial semiconductor foundries. While we have come a long way since the barium platinocyanide screen, today's scientific detectors use relatively old integrated circuit technology nodes. For example, NVIDIA's A100 graphics processing unit (GPU) is fabricated  in \SI{7}{\nano \meter} process introduced in 2018, while the  PSI/Dectris Eiger~\cite{Eiger1} X-ray detector ASIC is fabricated in  \SI{0.25}{\micro \meter} introduced in 1996. Detector technology has a long journey ahead until it reaches the end of Moore's law and consequentially one can expect larger and faster detectors to be fabricated for many years to come.

The general trend of making larger and faster detectors is imposing a huge burden on data storage, and more importantly, on data streaming. The latest generation of X-ray detectors \cite{Leonarski2020} boasts a raw data-rate of \SI{400}{Gbps}, more than one order of magnitude higher than the speed of a single laser-optimized multimode fiber. Thankfully, with the use of more advanced technology nodes, more digital functionality from the computing domains can be integrated directly into the detector ASIC. Among these functionalities, hardware-based streaming compression is finding its way into the modern computing~\cite{IBM-POWER9} and networking~\cite{Bluefield} ecosystem. 

In this paper, we describe an approach to increase detector frame rates using on-chip compression to reduce the amount of streaming data from a pixel array prior to being transmitted off-chip.  This architecture is scalable based on various design choices such as the maximum frame rate, ASIC technology, pixel array size, number of bits per pixel, etc. The goal is to support detector ASICs with a frame rates in the \SI{100}{\kHz} to \SI{1}{\MHz} regime. 

\section{Architectural Overview}

\begin{figure}[h]
    \centering
    \includegraphics[width=1.0\textwidth]{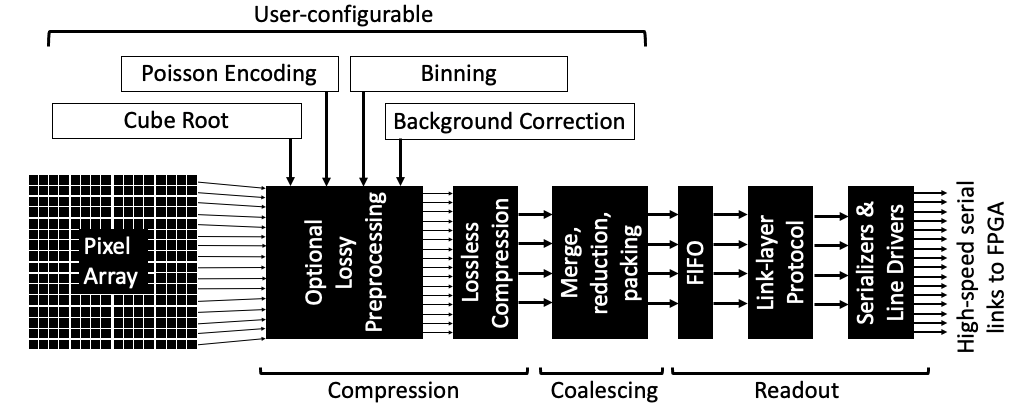}
    \caption{Architecture block diagram of a lightweight, user-configurable digital ASIC with on-chip data compression.}
    \label{fig:architecture_block_diagram}
\end{figure}

Figure~\ref{fig:architecture_block_diagram} shows a block diagram of our proposed detector ASIC architecture. The pixel array continuously shifts digital pixel data to the ASIC's edge, or "balcony", where the pixel data is digitally processed prior to being transmitted off-chip. The rows of the array act as parallel shift registers, continuously streaming pixel data from the array to the balcony for processing. An entire frame's worth of pixel data is shifted out of the array in one frame interval or less to support continuous readout. Multiple shift buses can be implemented per pixel row to reduce the time needed to shift all frame data to the edge and increase the possible frame rate.  
 
At the balcony, the shifting pixel data first passes through a preprocessing block which offers the detector user several methods for modifying the data prior to compression. Each preprocessing method can be independently enabled or disabled by the user and offers the potential to optimize the data compression results depending on the needs of the experiment, at the expense of some loss of information. Multiple methods can be used simultaneously.
 
After preprocessing, the pixel data passes through a lossless compression algorithm. This block leverages the fact that in a coherent diffraction imaging experiment the majority of the detector pixels measure zero or low counts most of the time. The method presented here allows low count pixels to be efficiently compressed.
 
Next the compressed data passes through the coalescing stage where the data is formatted and packed into wide words to be written into an asynchronous First In, First Out  (FIFO) memory in the readout stage. The FIFO absorbs temporary spikes and fluctuations in the bandwidth of the compressed data, and it bridges the variable bandwidth of the compressed data with the fixed bandwidth of the high-speed serial links. In the readout section, data from the FIFO is sent to a link layer block which prepares the compressed data to be transmitted over multiple high-speed serial links. Typical designs will have on the order of 16 serial links transmitting between \SI{1}{Gbps} to \SI{5}{Gbps} each.
 
\section{Compression Modules}
Compression reduces the size of the data by identifying and eliminating statistical redundancies. The input data characteristics affect both the complexity of the design and the compression ratio. In general, complex designs require significant effort to validate and test in an ASIC. Our goal is to identify a design that fulfills three requirements. First, it must yield reasonably high compression ratios for typical X-ray ptychography. Second, the operation must be stall-free; it must process input pixel data shifted from the pixel array every single clock cycle without causing any stall clock cycle or dropping any data. Finally, the design should also be simple and require minimal area on the detector ASIC. 
Based on these constraints, off the shelf compression algorithms (e.g., LZ4, LZ77, Huffman coding) have proven impractical. An ASIC implementation of LZ4~\cite{LeeLZ4} in \SI{65}{\nm} has a throughput rate of \SI{4}{Gbps}, which is two orders of magnitude slower than cutting-edge X-ray detectors \cite{Leonarski2020}. Instead we propose a lightweight lossless zero-suppression based compression in conjunction with a user configurable lossy data preprocessing stage, fine tuned to the statistical nature of a coherent diffraction imaging experiment. 

\subsection{Poisson encoding}

Poisson encoding refers to a lossy quantization scheme that exploits the statistical characteristics of Poisson (or shot) noise originating in the particle nature of the X-ray photons~\cite{Hammer_2021, Panpan}. The standard deviation of shot noise is $\sqrt{N}$ where $N$ is the number of photons (or events) hitting each pixel for the duration of the data acquisition. Because the uncertainty of the measurement scales with the square root of the photon counts, so does the step size of the quantization. The goal is to maximize the efficiency of the lossy compression algorithm while having equal and minimum impact on all the detector pixels regardless of their photon counts.

\begin{figure}[h]
    \centering
    \includegraphics[width=0.95\textwidth]{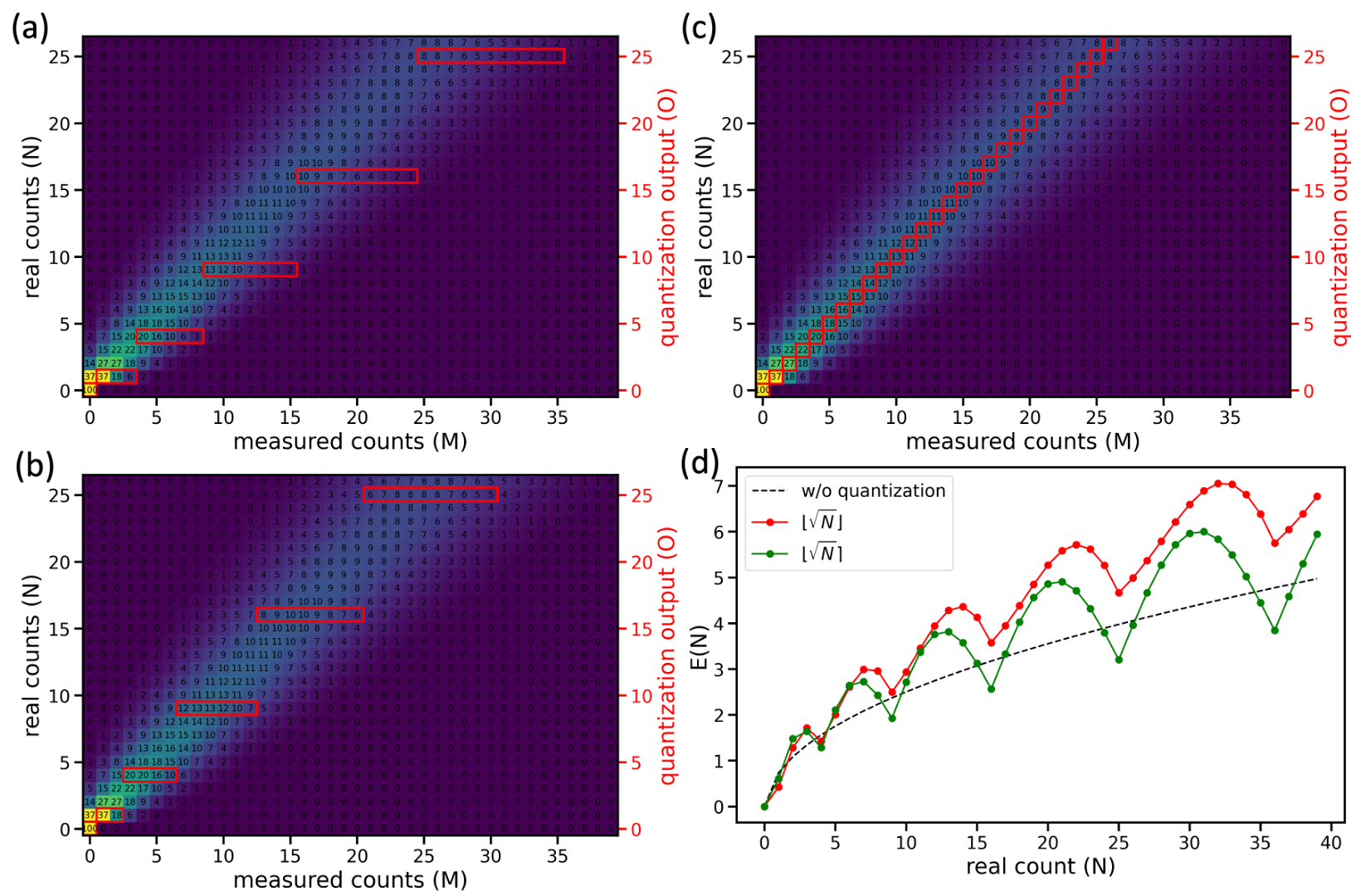}
    \caption{Different implementations of Poisson encoding with (a) $R=\lfloor\sqrt{N}\rfloor$, (b) $R=\lfloor\sqrt{N}\rceil$ and (c) without quantization. The probability for a real photon count $N$ to be measured as $M$ is shown in fake color scale and annotated in each cell. The sum of each row is thus $100\%$. The output of the quantization is highlighted by the red rectangle. (d) A comparison of their mean error $E(N)$ as a function of measured counts $N$.}
    \label{fig:poisson-compression}
\end{figure}

The simplest way to implement Poisson encoding is to take $R=\lfloor\sqrt{N}\rfloor$, where $\lfloor\rfloor$ denotes the floor operation. The decoded value of the quantization output is $O=R^2$. On an ASIC this can be implemented efficiently by iteratively computing the binary digits of the square root result~\cite{ming2011fundamental}. The accuracy of such an implementation is shown in Figure~\ref{fig:poisson-compression}a. It can be seen that the output of the quantization is not centered on the probability mass function (pmf) of the Poisson distribution. For instance, a measured count of 15 is quantized as 9 counts, but there is only a $1.9\%$ chance for a real photon count of $N=9$ to be measured as $M=15$. Being off-centered on the pmf also leads to a larger mean error between the quantization output $O$ and the real counts $N$, calculated as $E(N) = \int|N-O(M)| \times p(M,N) dM$ where $p(M,N)$ is the probability mass function. For $O(15) = 9$ we have $E(15) = 4.13$. This error metric is particularly relevant for numerical optimization scheme such as Gerchberg-Saxton type~\cite{Fienup1982, Nashed2014} or gradient based~\cite{Kandel2019, Zhou2021} iterative phase retrieval. The error of the quantization can be greatly reduced by using $\lfloor\sqrt{N}\rceil$, where $\lfloor\rceil$ denotes the rounding operation. This time (Figure~\ref{fig:poisson-compression}b), with $O(15) = 16$, the mean error $E(15)$ is just 3.13. For comparison, Figure~\ref{fig:poisson-compression}c shows the output when Poisson encoding is disabled (i.e. without quantization).

Figure~\ref{fig:poisson-compression}d shows the mean error $E(N)$ for different Poisson encoding schemes as a function of real counts $N$. The dashed line shows the baseline without quantization. While Poisson encoding with rounding operation has a lower mean error for $N\ge7$, the simple implementation with the floor operation actually performs better at the low photon counting regime, in particular for $N=1$ and $N=2$. As will be shown later, this has potential impact on the preservation of spatial resolution because high frequency information is often carried by low counts pixels in a typical coherent diffraction imaging experiment. Interestingly, the raw counts are not always more accurate than with lossy Poisson encoding, in particular near $N = 1^2, 2^2, 3^2...$. This is because an actual count of $N=16$ can be measured as any number of counts such as $M=13$ or, albeit far less likely, $M=1$ . The Poisson encoding with rounding operation in this case quantizes everything between $M\in[13,20]$ to $O(M)=16$. This increases the probability of recovering an actual count of $N=16$ and in turn reduces its mean error at $E(16)$.

Both Poisson encoding schemes $R=\lfloor\sqrt{N}\rfloor$ and $R=\lfloor\sqrt{N}\rceil$ are available and can be switched depending on the preference of the user. Algorithm~\ref{alg:sqrt} shows an example that can be efficiently implemented in ASIC. Through the use of helper variables when calculating $\lfloor\sqrt{N}\rfloor$, this method does not require any expensive multiplication or division steps which significantly reduces the resource cost. The optional rounding is achieved by checking if calculating $R$ to the first binary digit after the radix point would make $R^2$ exceed $N$. A single-cycle 10-bit square root module uses 494 CMOS cells (and, mux, not, or, xor) and occupies an area of $\SI{20}{\um} \times \SI{20}{\um}$ in \SI{130}{\nm} CMOS. 
Since the addition of 64 of such Poisson encoding modules increases the area of the circuit (including the lossless compression module and the coalescing module) only by $30\%$, we are confident that the proposed implementation is feasible in a \SI{130}{\nm} design.

\begin{algorithm}
\small
\caption{Calculate the square root $S=\sqrt{N}$ with optional rounding controlled by the $R$ input using unsigned integer math}\label{alg:sqrt}
\begin{algorithmic}[1]
\Require $0 \leq N < 2^b$, $R\in\{\textit{false}, \textit{true}\}$
\State $m \gets \text{ceil}(b/2)$ \Comment{The number of bits needed to represent the square root $R$}
\State $C_{m-1} \gets 0$ \Comment{Helper variable $C_{n-1}=2^{n}R_{n}$}
\State $D_m \gets N$ \Comment{The difference $N-R_n^2$}
\For{$n \gets m - 1, \ldots, 2, 1, 0$} \Comment{$n$ is the index of the output bit to check}
\If{$D_{n+1} \geq C_n + 2^{2n}$}
\State $D_n \gets D_{n+1} - C_n - 2^{2n}$
\State $C_{n-1} \gets (C_n \gg 1) + 2^{2n}$ \Comment{$\gg$ denotes the bit-shift right operation}
\Else
\State $D_n \gets D_{n+1}$
\State $C_{n-1} \gets C_n \gg 1$
\EndIf
\EndFor
\If{$R \; \textbf{and} \; D_0 > C_{-1}$} \Comment{Round the square root if $R=\textit{true}$}
\State $S \gets C_{-1} + 1$
\Else
\State $S \gets C_{-1}$
\EndIf
\end{algorithmic}
\end{algorithm}

\subsection{Parasitic background correction, binning and cropping}
For a detector with near $100\%$ quantum efficiency (which is usually true for the latest generations of pixel detectors working at their optimal energy range), the accuracy of Poisson encoding relies on the shot noise nature of the counting statistics. When the noise distribution is not purely Poissonian, for instance, if strong scattering from air or the sample environment is present during the experiment, the simplest fix is to subtract this background noise (called parasitic background correction) before applying Poisson encoding. Such correction can improve not only the accuracy of the Poisson encoding scheme, but also the compression ratio of the data.

When an even higher data transmission rate is desired, binning of the neighbouring pixels and cropping of the edge of the detector can be applied. For coherent diffraction imaging methods, binning preserves high frequency information but reduces the oversampling ratio of the data in the Fourier space while cropping retains the same oversampling ratio at the potential cost of a reduced spatial resolution.

\subsection{Bit shuffling}
Bit shuffling~\cite{Bitshuffle} works by rearranging the bits in an array of pixels so that the bits at the same position in the binary representation of each pixel value are grouped together. An illustration of the process can be found on the left hand side of Figure~\ref{fig:dbw-compression}. Though not inherently a compression technique, bit shuffling maximizes the number of zero-valued bit "groups" in low-count pixel regimes, which subsequent zero-suppression compression schemes can exploit. It is important to note that bit shuffling does not require any logic and can be implemented at zero cost by simply rearranging wires in the circuit.

\subsection{Dynamic bit-width zero suppression}
The most common redundancy in X-ray diffraction data are the leading zeros in low counts pixels. Most X-ray data consist of only small highly illuminated regions, with the rest of the pixels seeing low or zero photon counts. A lightweight and effective way to reduce the amount of data using a form of zero-suppression is to dynamically adjust the bit-width of pixels to match their values, using the minimum number of bits necessary to represent them and therefore transmitting less data. Doing this for every pixel independently is not feasible since there needs to be some way to transmit how many bits are used to represent each pixel. In conjunction with bit shuffling, this method can be applied to small pixel regions by bit-shuffling their pixels and working on the output, which means using the same number of bits to represent all pixels in a region. The compression output for each small pixel region becomes a variable number of words (up to the original number of bits in a pixel) and a number indicating how many words, or bits per pixel, there are. This method avoids relying on any patterns in the X-ray data besides assuming that neighboring pixels roughly have the same value and therefore bit-width. 

\begin{figure}[h]
    \centering
    \includegraphics[width=0.75\textwidth]{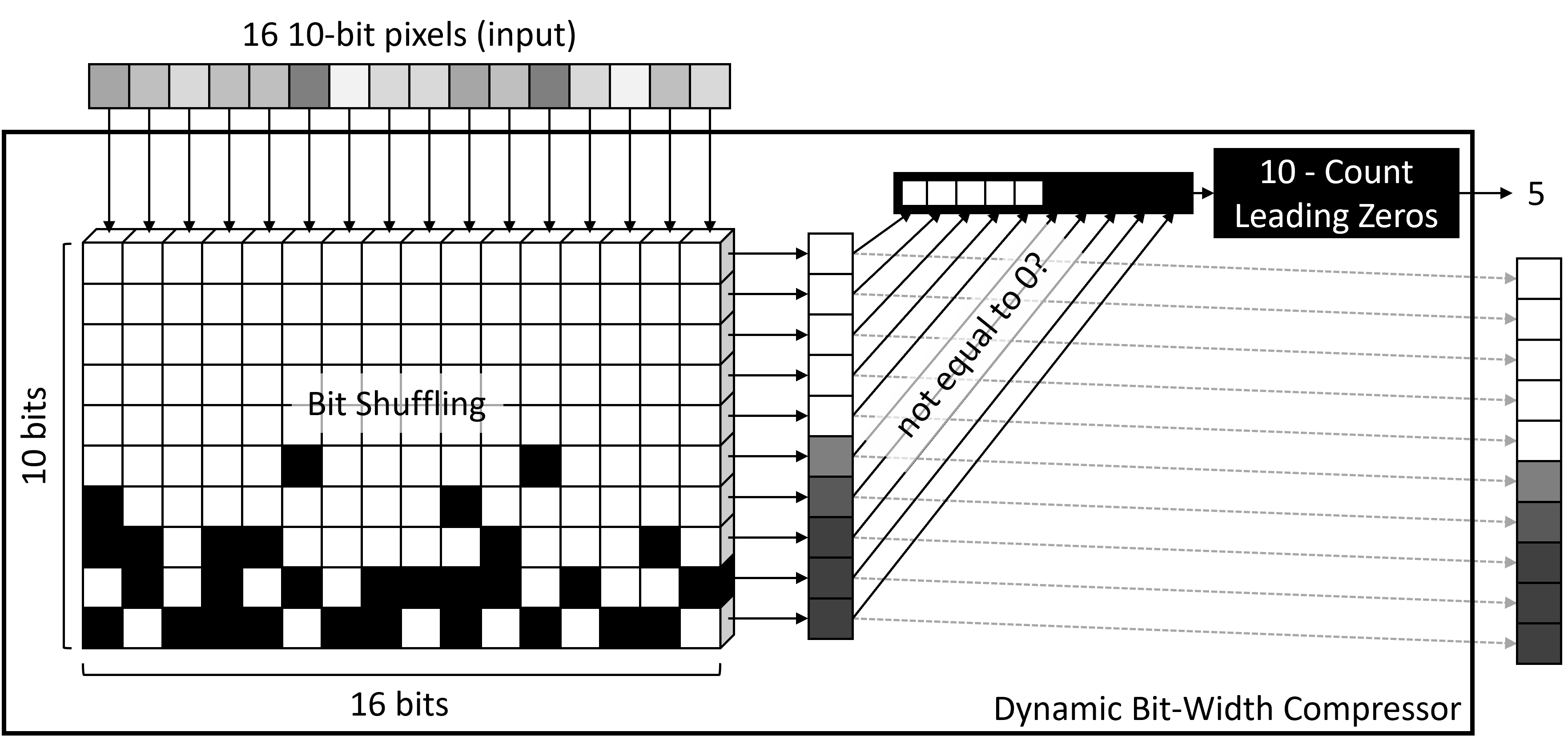}
    \caption{Overview of a Dynamic Bit-Width Compression module which processes an area of 16 10-bit pixels and outputs the bit-shuffled data and the number of continuous nonzero elements (the bit-width). In this example, all input pixels have values that can be represented in at most 5 bits. Therefore only the outputs representing the 5 least significant bits of each pixel contain information.}
    \label{fig:dbw-compression}
\end{figure}

The main advantage of this compression method over more common lossless algorithms such as LZ4 is its low resource usage while still achieving a decent compression ratio on typical X-ray ptychography data~\cite{Hammer_2021}. To determine the number of bits needed to represent a value, simple count leading zeros logic is used. Since multiple compressors are running in parallel to compress each pixel region, the most resource-intensive part is combining their output into one continuous data stream. That logic is further discussed in section~\ref{coalescing}. A rough schematic of a compression module for a 4 $\times$ 4 pixel region is shown in Figure~\ref{fig:dbw-compression}.

\subsection{Performance of the compression modules}

\begin{figure}[h]
    \centering
    \includegraphics[width=0.95\textwidth]{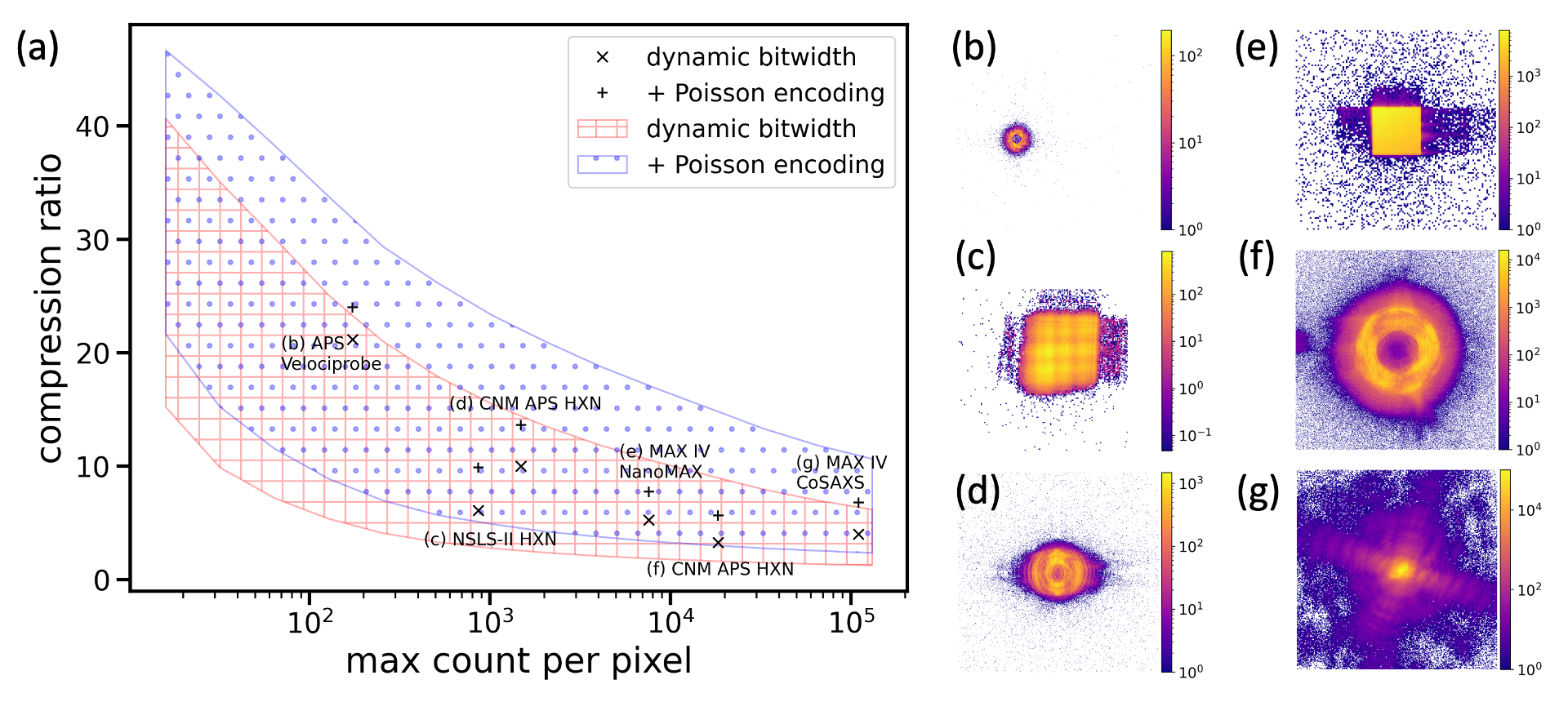}
    \caption{(a) Estimated range of the compression ratio as a function of max counts per pixel, for pure lossless compression and for adding Poisson encoding. Compression ratio calculated on experimental datasets acquired at beamlines across multiple synchrotron facilities were also shown. (b)-(f) Example detector image for the forward ptychography experimental datasets used in the calculation. (g) Example detector image for the tele-ptychography experimental datasets used in the calculation.}
    \label{fig:comp-comparison}
\end{figure}

The effectiveness of the compression module was first estimated by rescaling and resizing a high count rate experimental dataset~\cite{Anakha-edge-inference}. Rescaling the detector intensity and adding Poisson noise allows several artificial datasets to be created with maximum count per pixel ranging from 15 (4 bits) to 131071 (17 bits). Resizing of the detector image generates additional datasets with different magnifications of the transmitted beam, which accounts for, for instance, variations in the sample to detector distances at different beamlines. The result is shown in Figure~\ref{fig:comp-comparison}. With pure lossless algorithms (i.e. bit shuffling plus dynamic bit-width zero suppression), the compression ratio can be as low as $1.2\times$ for high count rate data with also a high magnification of the transmitted beam or as high as $40\times$ in the exact opposite case. Adding Poisson encoding ($R=\lfloor\sqrt{N}\rfloor$ or $R=\lfloor\sqrt{N}\rceil$) typically results in a flat increase of on average 3 in the compression ratio, regardless of the count rate and magnification of the transmitted beam. 

The estimated compression values were further supported by results calculated on a variety of experimental datasets acquired at beamlines across multiple synchrotron facilities ~\cite{Anakha-edge-inference,Kahnt:dy5004,Huang:2017ca, Jiang:2021jw}. An example of the detector images for each of these data are shown in Figure~\ref{fig:comp-comparison}b-g. All the experimental compression ratios fall well within the range of the estimation. The flat increase of roughly 3 in compression ratio after introducing Poisson encoding was also observed.

\begin{table}[ht]
\small
\centering
\begin{tabular}{|l|r|r|r|r|r|} 
 \hline
 \textbf{Dataset Name} & \textbf{dynamic bitwidth} & \textbf{+ Poisson} & \textbf{PNG} & \textbf{JP2 lossless} & \textbf{JP2 lossy q=90} \\ 
 \hline
 CNM APS HXN \#1~\cite{Anakha-edge-inference} & 9.98   & 13.62  & 8.45	& 12.98	& 20.31 \\ 
 \hline
 CNM APS HXN \#2~\cite{Anakha-edge-inference} & 3.27	& 5.66	& 2.56	& 3.42	& 6.68 \\
 \hline
 MAX IV NanoMAX~\cite{Kahnt:dy5004} & 5.26	& 7.74	& 4.16	& 5.18	& 19.22 \\
 \hline
 NSLS-II HXN~\cite{Huang:2017ca} & 6.08	& 9.89	& 5.03	& 7.00	& 18.92 \\
 \hline
 APS Velociprobe ~\cite{Jiang:2021jw} & 21.16	& 24.02	& 31.66	& 54.55	& 792.35 \\
 \hline
\end{tabular}
\caption{Table of compression ratios with a selection of preprocessing methods and datasets.}
\label{tbl:ratios}
\end{table}
For comparison purposes, Table~\ref{tbl:ratios} shows the compression ratios obtained from standard lossless (portable network graphics or PNG~\cite{boutell1997png}) and lossy (JPEG 2000 or JP2~\cite{taubman2012jpeg2000}) image compression algorithms on the experimental datasets. PNG is an image file format which supports lossless compression using a two-stage image compression process that consists of filtering (prediction) and DEFLATE, which is a combination of LZ77~\cite{ziv1977universal} and Huffman coding~\cite{huffman1952method}. JP2 transforms images with discrete wavelet transform (DWT), quantizes DWT coefficients, and then encodes them using embedded block coding with optimal truncation (EBCOT)~\cite{taubman2000high}. JP2 compression was tested without quantization (i.e. lossless) and with a quality factor of 90. In general, the compression ratio for our dynamic bit-width compression with bit shuffling performs slightly better than PNG but slightly worse than JP2 without quantization, all three being lossless compression schemes. For lossy compression, JP2 fares much better than Poisson encoding even at a high quality factor of 90. It is worth nothing that with JP2 lossy compression all pixels suffers indiscriminate loss while with Poisson encoding the loss is physics-aware and tailored to the statistical nature of X-ray science.

\subsection{Impact of Poisson encoding}

\begin{figure}[h]
    \centering
    \includegraphics[width=0.9\textwidth]{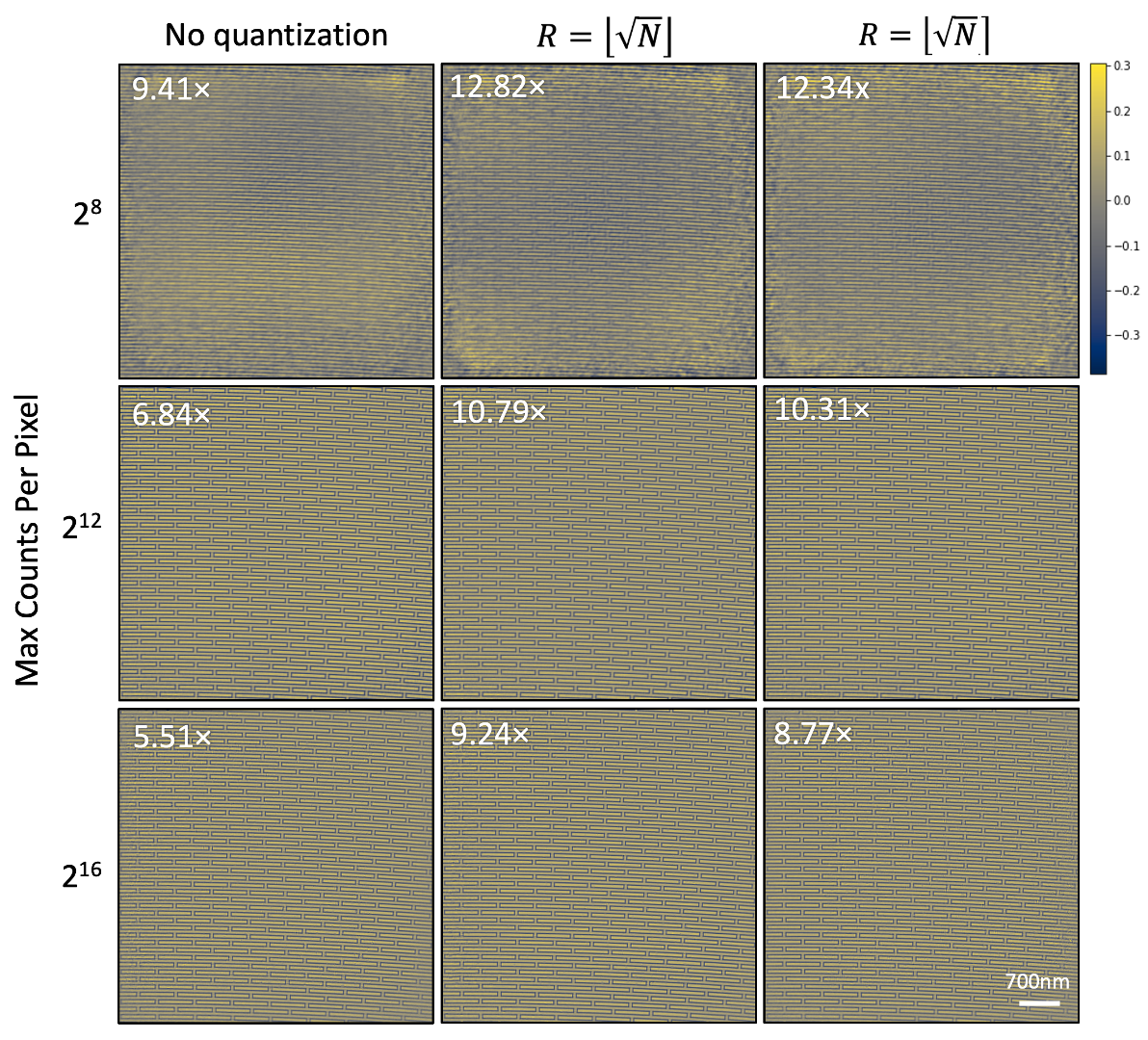}
    \caption{Reconstruction results on simulated ptychography datasets with 14 nm feature size. Data with various count rates and added Poisson noise were used. The number at the top left corner of each image denotes the compression ratio.}
    \label{fig:comp-effects}
\end{figure}

The impact of Poisson encoding on the reconstruction quality of X-ray ptychography was evaluated on a simulated data set with \SI{14}{\nano \meter} feature size. We chose specifically this dataset over other published experimental data because of its high sensitivity to the loss of information and its high spatial resolution. Using a simulated dataset also allowed us to rescale the per pixel intensity at will and test the impact of different count rates on the reconstructed image. The result is shown in Figure~\ref{fig:comp-effects}. For medium to high count rates (max counts per pixel $>2^{10}$), the reconstruction on data with lossy Poisson encoding performed equally well compared to those compressed only losslessly (i.e. without quantization). For the dataset with max counts per pixel of $2^8$, the data without quantization showed visible loss of the spatial resolution while the data with $R=\lfloor\sqrt{N}\rfloor$ showed the highest reconstruction quality. This might be puzzling at first glance considering that the floor operation should result in, in principle, the most information loss. However, it can be easily understood by reviewing the error function shown in Figure~\ref{fig:poisson-compression}d. The Poisson encoding with floor operation has the lowest error for all the pixels with real counts $N=1$. At high count rates, those pixels are mostly found at the peripheral of the diffraction patterns, corresponding to areas carrying only the highest spatial frequency information. As the count rate decreases, those pixels become increasingly important as they start to dominate the mid-to-low spatial frequency regime. This extra benefit justifies why, despite its higher information loss at $N\ge7$, the Poisson encoding scheme with the floor operation was implemented in our design in addition to the more generally accepted scheme with the rounding operation. It is worth noting that such effect may be too subtle to be noticed in experimental data due to influence from probe coherence, positional error, background noise, etc.

\section{Coalescing modules}\label{coalescing}

\begin{figure}[h]
    \centering
    \includegraphics[width=0.75\textwidth]{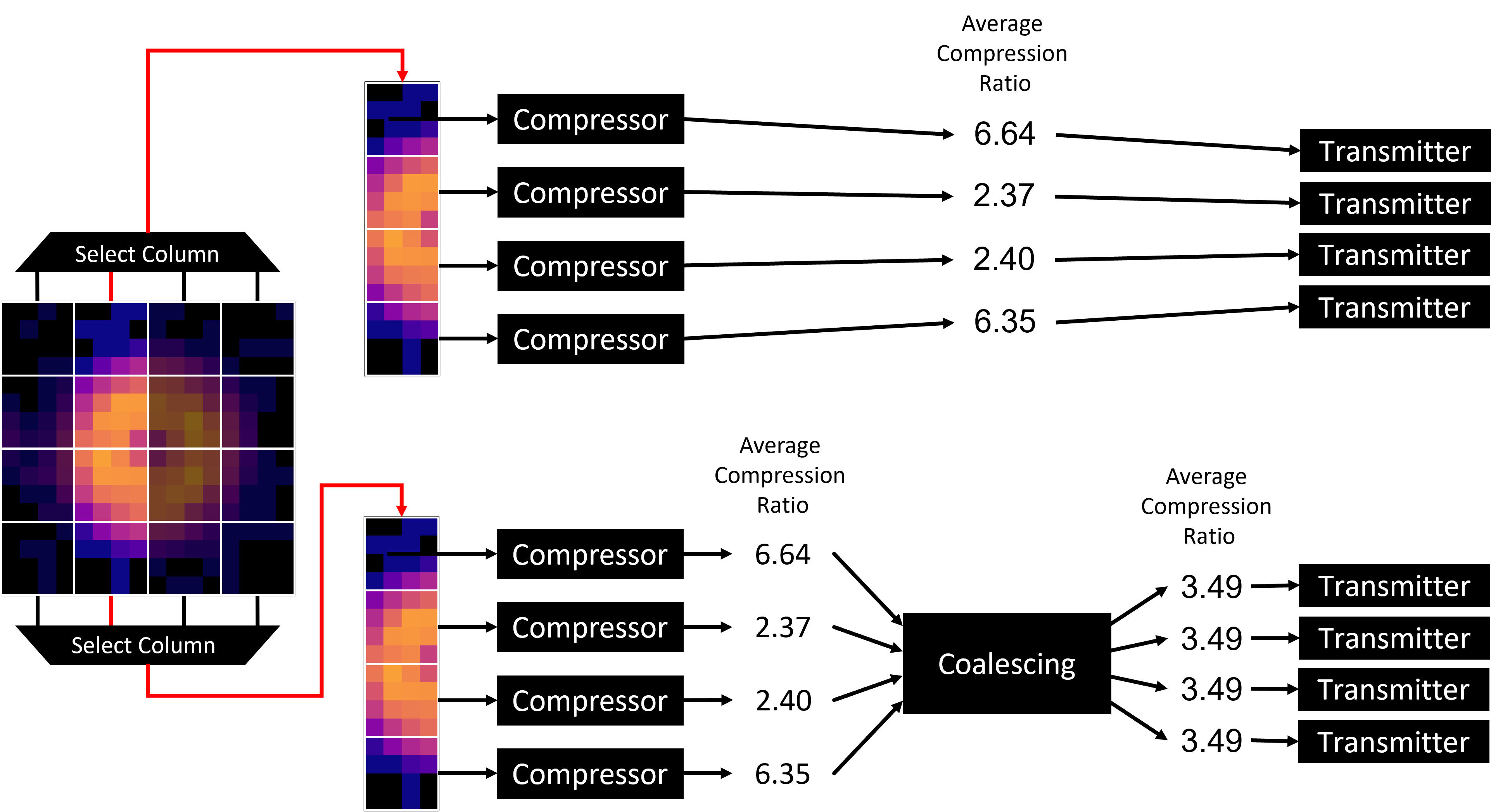}
    \caption{Example mapping of pixels to the compressors for a 16 $\times$ 16 pixel array with 4 compressors with and without coalescing. The full image is compressed over four clock cycles column by column. The output of the compressors is combined by the coalescing module to distribute the data between the transmitters. Without the coalescing module the data would not be evenly distributed between the transmitters, resulting in bottlenecks.}
    \label{fig:comp-mapping}
\end{figure}

Since the data transmission rate is limited by the compressor block with the lowest compression ratio, it is critical that the output data from the compressor blocks are distributed evenly together across all the parallel transmitters. An example pixel to compressor mapping is shown in Figure~\ref{fig:comp-mapping}. It can be seen that compressors assigned to rows containing the high photon count regions generate considerably more data due to their low compression ratio. To address this, a coalescing module which combines the variable-length data from the compressors and outputs fixed sized words is used~\cite{Strempfer:coalescing} to evenly distribute the data across the transmitters.

\begin{figure}[h]
    \centering
    \includegraphics[width=0.5\textwidth]{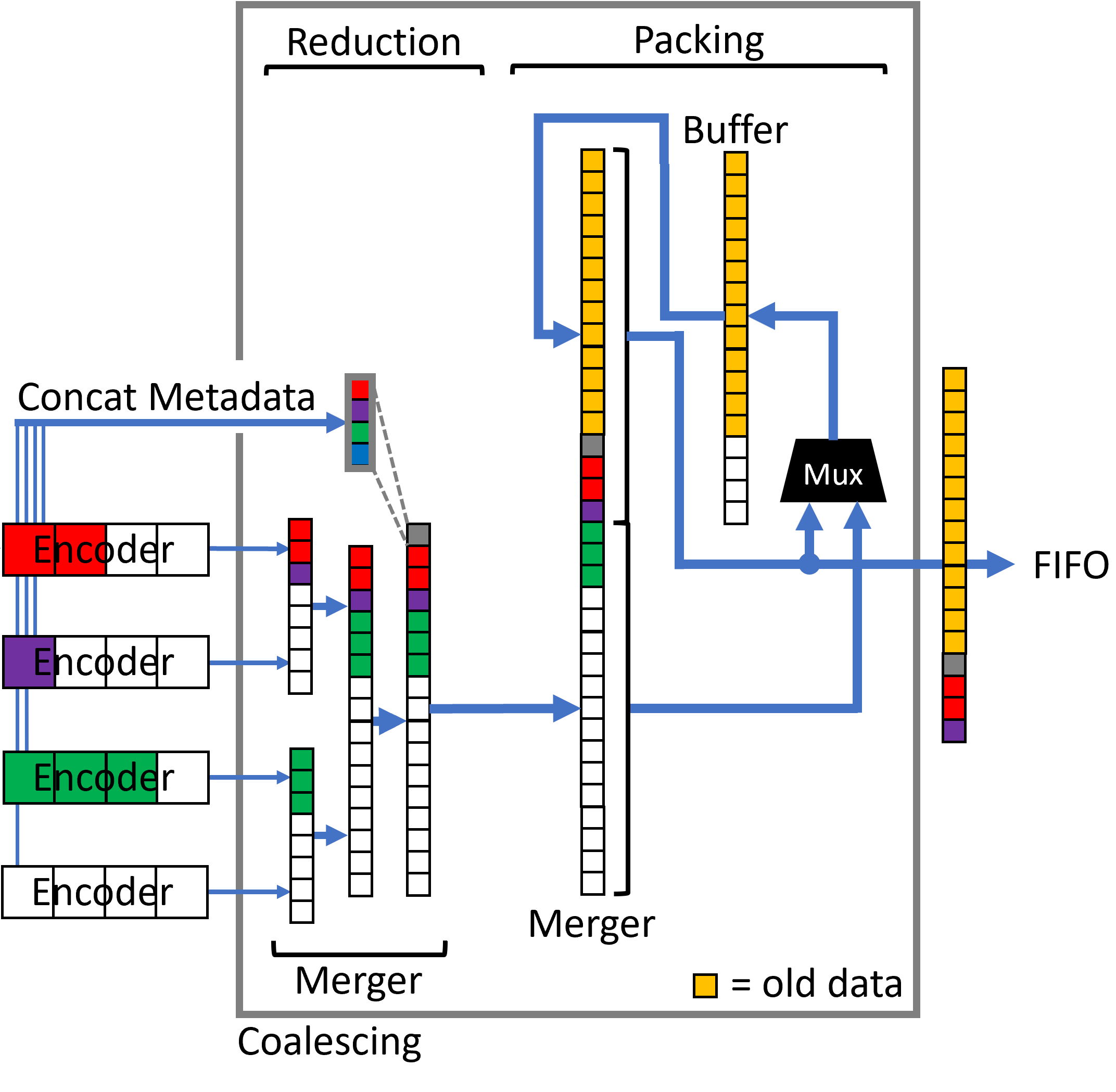}
    \caption{Conceptual layout of coalescing logic for four parallel encoders/compressors with example data. The encoder output is merged to be continuous and then packed into constant-sized words of data. In this example, the green data will be stored in the Buffer during the next clock cycle.}
    \label{fig:comp-overview}
\end{figure}

The main challenge of designing such a data packing mechanism is their higher resource usage. For example, the variable to fixed length converter in a bandwidth compressor design consumes more than $80\%$ of the entire resources~\cite{Ueno:2017hv}. Also challenging is the performance optimization of such implementations, particularly for high-speed designs and with field-programmable gate arrays (FPGAs). Our solution is to use a data coalescing module composed of two distinct stages: reduction and packing (Figure~\ref{fig:comp-overview}). The reduction stage takes the output of multiple parallel encoders, consisting of both constant and variable-length data and turns it into a single continuous variable-length output. The implementation and resource optimization strategy is described in our previous work~\cite{yoshii2021hardware}. The packing module then combines this variable-length data across clock cycles until a fixed number of bits has accumulated. This constant-sized data is then passed on to the FIFO buffer, to be transmitted off-chip when ready. Aside from the buffer in the packing module, the coalescing logic is entirely combinational. This choice simplifies the digital logic design since it allows the compressor to use the same clock frequency as the rest of the system. 

\section{Readout Modules}

Custom logic design and common ASIC IP components (differential serial line drivers, phase-locked loops, open source data protocols, FIFOs, SRAM blocks) can be used to build a high-speed ASIC interface which can buffer the data from a pixel array and transmit it off-chip to an FPGA. This section will discuss an example readout block which allows variable rate compressed data to be transmitted off-chip over fixed bandwidth serial channels.

\begin{figure}[h]
    \centering
    \includegraphics[width=1.0\textwidth]{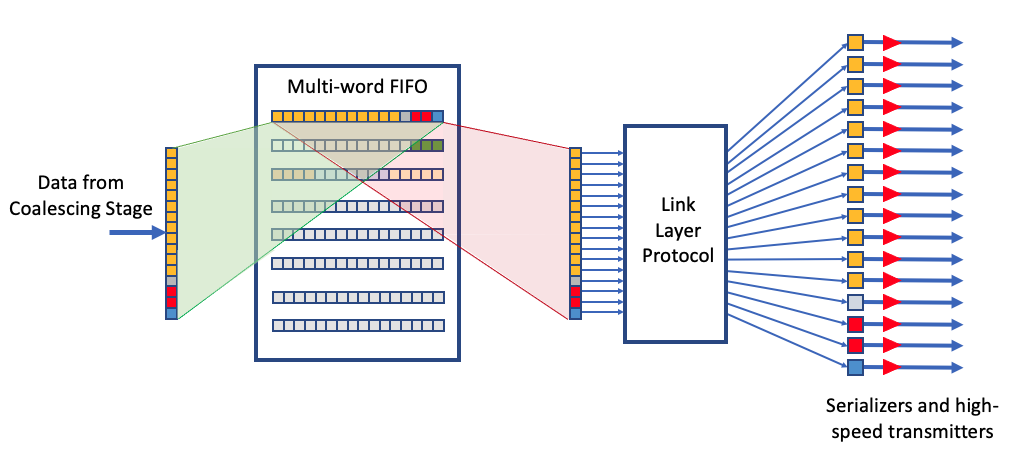}
    \caption{Block diagram of an example readout section which includes a wide asynchronous FIFO and a Link Layer Protocol block. This section absorbs peaks in the compressed data which temporarily exceed the off-chip transmit bandwidth.}
    \label{fig:large_fifo}
\end{figure}

The aforementioned coalescing logic can only balance variations in output data size across parallel compressors. Another challenge with on-chip compression is how to handle the variable data velocity from fluctuations in raw data. The intensity and distribution of scattered X-ray may change during an experimental scan. There must be a mechanism to absorb these fluctuations before the pixel data can be transmitted. We propose a readout section composed of a large asynchronous FIFO with a standardized link layer protocol driving multiple high-speed serializers, as shown in Figure~\ref{fig:large_fifo}.
 
\subsection{FIFO}
A wide asynchronous FIFO is useful in this architecture since the FIFO can be sized such that the word width matches exactly the number of bits needed to fill all of the serializers for each clock cycle. This simplifies the read side of the FIFO as each read operation can directly load all serializers. On the FIFO write side, a wide word width allows the coalescing stage to efficiently pack each FIFO word, accumulating merged data over multiple clock cycles before writing. A wide asynchronous FIFO serves several purposes:

\begin{enumerate}
\item Matches asynchronous read and write clock domains, relaxing clock constraints and clock routing restrictions
\item Accepts wide packed data words accumulated by the coalescing stage. This allows the coalescing stage to pack more efficiently
\item Evens out peaks and valleys in frame data by allowing the FIFO to fill during intervals of peak data, and spooling that data into the fixed bandwidth of the high-speed transmitters during clock cycles with lesser pixel data
\item Readily loads the high-speed output serializers each clock cycle, minimizing IDLE cycles
\end{enumerate}

\subsection{Link layer protocol}
Even with compression, multiple high-speed serial transmitters operating in parallel are needed to provide the off-chip bandwidth necessary for a \SI{100}{\kHz} to \SI{1}{\MHz} frame rate. As pixel data is transmitted, some way of maintaining frame integrity across parallel serial data streams is needed. Ideally the compressed data is transmitted within a protocol layer which bonds the serial lanes together and has the following properties:

\begin{enumerate}
\item Is well documented and easily accessible
\item Allows for a lightweight implementation by supporting simplex-only transmission
\item For simplicity does not require clock compensation, but can support it if required
\item Does not require flow control. The ASIC design is simpler if the FIFO is sized only to absorb peaks in the out-going compressed data, and does not also need to compensate for the overflowing of downstream buffers.
\item Provides readily available IP blocks for the link layer interface in a receiving FPGA
\item Supports channel bonding so multiple serial transmit lanes can be word synchronized in a receiving FPGA
\item Supports messaging over the interface with block codes such as IDLE to indicate an empty transmit FIFO, or codes to signal other user-defined messages such as FIFO Overflow
\item Uses a standardized line encoding format
\end{enumerate}

A Xilinx Aurora interface~\cite{aurora} matches these requirements well. Other similar protocols can also be used. 

\section{Implementation in \SI{130}{\nm} CMOS}

We include ASIC implementation (i.e., synthesis and place-and-route) results for several blocks which would be needed for a small array of 64 $\times$ 64 10-bit pixels in \SI{130}{\nm} CMOS. Other configurations with larger arrays are achievable as the digital logic of the architecture described will scale with smaller CMOS processes. \SI{130}{\nm} CMOS was selected for our example design because it allows relatively low prototyping cost and simple physical design. A relatively small 64 $\times$ 64 pixel array was chosen to study the implementation of a potential detector test chip.

\begin{figure}[h]
    \centering
    \includegraphics[width=1.0\textwidth]{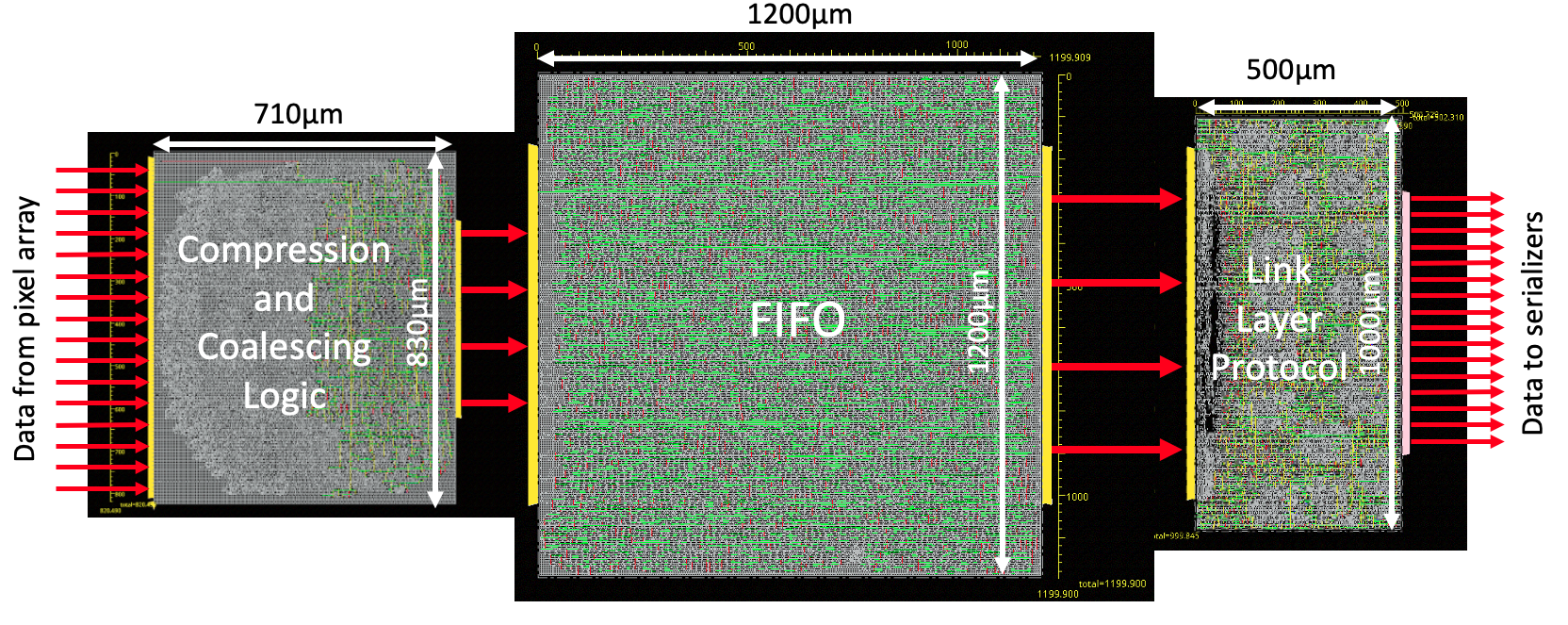}
    \caption{Example layouts of several of the blocks in \SI{130}{\nm} CMOS.}
    \label{fig:layout}
\end{figure}

An example layout of the compression and coalescing block discussed earlier is shown in Figure~\ref{fig:layout} and occupies 0.6 mm$^2$. This example block accepts 64 10-bit pixels every clock cycle and outputs one 1024-bit word to a FIFO once it has accumulated enough data. Inputs are shown on the left and outputs on the right, both in yellow. The light grey "cloud" in the compression block layout are the individual logic standard cells viewed from a high level. The green routing of the clock net, which is almost exclusively on the right half, demonstrates that the combinational logic of the compression and reduction circuits on the left (input) side of the block are not clocked. The clocked registers of the packing circuits which drive the outputs are on the right.

Figure~\ref{fig:layout} also shows the layout of an example FIFO containing 16 1024-bit words which, for simplicity, uses registers as the storage elements. This example 16 Kbit FIFO in \SI{130}{\nm} CMOS occupies an area of 1.4 mm$^2$. The low number of storage bits in this example approaches the smallest usable FIFO size for a 64 $\times$ 64 pixel array while replacing registers with SRAM would reduce the area and allow for a larger FIFO. In practice this FIFO can be sized up to fill whatever chip area could be allocated to it.

The layout of a 16-lane Aurora interface circuit which provides the link-layer protocol is also shown in Figure~\ref{fig:layout}. This block includes a FIFO read controller, 16 64B/66B scramblers, 16 gearboxes, and drives output data to 16 32-bit serializers (not included in the layout). It also inserts standard Aurora IDLE and Channel Bonding block codes under external control.

For a 64 $\times$ 64 array of \SI{100}{\um} pixels (41 mm$^2$) implemented in \SI{130}{\nm} CMOS, the total area for the previously described circuits (compression, coalescing, and readout) could be limited to 4 mm$^2$ to fit in a balcony area that is less than 10\% of the area of the pixel matrix.

\section{Conclusion}
We have described a lightweight, user-configurable detector ASIC digital architecture with on-chip compression for X-ray coherent diffraction imaging experiments with a frame rate in the \SI{100}{\kHz} to \SI{1}{\MHz} regime. Synthesis and place-and-route results of the major blocks (compression, coalescing and readout) are provided as proof of feasibility for a small array of 64 $\times$ 64 10-bit pixels in \SI{130}{\nm} CMOS. Both the detector size and the technology nodes are chosen for reasons of low prototyping cost and simple physical design, but the proposed architecture is scalable to a larger pixel array.

Even for a small array of 64 $\times$ 64 10-bit pixels, running at \SI{1}{\MHz} would produce a raw data rate in excess of \SI{40}{Gbps} that would be challenging to be transmitted off-chip without any compression. The proposed architecture includes both a user-configurable lossy preprocessing stage and a lightweight lossless compression stage, The lossless compression stage leverages the characteristic intensity distribution of X-ray scattering while the Poisson encoding preprocessing exploits the statistical nature of shot noise variations. Together they offer a reliable compression of data that was validated using experimental datasets gathered across multiple synchrotron facilities. A ptychography beamline at the next generation diffraction limit X-ray source~\cite{APSU} typically has $10^{11} - 10^{12}$ photons per second going through their focusing optics. At the maximum frame rate of \SI{1}{\MHz}, we expect the compression ratio to be at least 10$\times$ based on estimations in Figure~\ref{fig:comp-comparison}. That reduces the data rate to \SI{4}{Gbps} which can then be transferred through 4$\times$ \SI{1.28}{Gbps} serial links. It is worth mentioning that our compression blocks offer not only a much higher throughput rate (10$\times$) comparing to the ASIC implementation of off-the-shelf LZ4~\cite{LeeLZ4} algorithm running at similar clock speeds, but also a higher compression ratio at a lower gate count (\textasciitilde 4$\times$). The proposed detector architecture is a practical solution for not only coherent imaging techniques such as ptychography, but also for other X-ray scattering based techniques running at extremely high frame rates.

\acknowledgments
We would like to thank Lorenzo Rota, Aseem Gupta, Dionisio Doering and Angelo Dragone for helpful discussions on detector ASIC design. We would like to thank Michael Wojcik for the sample designs used in for the simulated datasets. We thank Alexander Bjorling, Xiaojing Huang and Yi Jiang for access to experimental datasets. This work was supported by the Laboratory Directed Research and Development (2021-0072) program at Argonne National Laboratory. Work performed at the Center for Nanoscale Materials and Advanced Photon Source, both U.S. Department of Energy Office of Science User Facilities, was supported by the U.S. DOE, Office of Basic Energy Sciences, under Contract No. DE-AC02-06CH11357. This research also used the resources of the SLAC National Accelerator Laboratory. A.K and M.J.C also acknowledge support from the U.S. Department of Energy, Office of Science, Office of Basic Energy Sciences Data, Artificial Intelligence and Machine Learning at DOE Scientific User Facilities program under Award Number 34532.  

\bibliographystyle{ieeetr}
\bibliography{references}
\end{document}

%% file: license.tex
\section*{Government License}

The submitted manuscript has been created by UChicago Argonne, LLC,
Operator of Argonne National Laboratory (“Argonne”). Argonne, a
U.S. Department of Energy Office of Science laboratory, is operated
under Contract No. DE-AC02-06CH11357. The U.S. Government retains for
itself, and others acting on its behalf, a paid-up nonexclusive,
irrevocable worldwide license in said article to reproduce, prepare
derivative works, distribute copies to the public, and perform
publicly and display publicly, by or on behalf of the Government.  The
Department of Energy will provide public access to these results of
federally sponsored research in accordance with the DOE Public Access
Plan. http://energy.gov/downloads/doe-public-access-plan.

\newpage